\documentclass{aip-cp}

\usepackage[numbers]{natbib}
\usepackage{rotating}
\usepackage{graphicx}
\usepackage{url}

\begin{document}

\title{Excited Nucleon Spectrum and Structure Studies with CLAS and CLAS12}

\author{Daniel S. Carman\\[0.7ex] {\large (for the CLAS Collaboration)}}
\affil{Thomas Jefferson National Accelerator Facility, Newport News, VA 23606}
\corresp{email: carman@jlab.org}

\maketitle

\begin{abstract}
  The study of the spectrum and structure of excited nucleon states employing the electroproduction
  of exclusive reactions is an important avenue for exploring the nature of the non-perturbative strong
  interaction. The CLAS detector in Hall~B has provided the dominant part of the available world data
  on most relevant meson electroproduction channels off the nucleon in the resonance region for $Q^2$
  up to 5~GeV$^2$. Analyses of CLAS data for the exclusive channels $\pi N$, $\eta N$, and
  $\pi^+ \pi^- p$ on a proton target have provided the only results available on the $Q^2$ evolution of
  the electro-excitation amplitudes for the transitions from the initial photon-proton to the final $N^*$
  states in the mass range up to $W$=1.8~GeV. These electrocouplings allow for exploration of the
  internal structure of the produced excited nucleon states. This work has made it clear that consistent
  results from independent analyses of several exclusive channels with different resonance hadronic
  decay parameters and non-resonant backgrounds but the same $N^*$ electro-excitation amplitudes, is
  essential to have confidence in the extracted results. Starting in early 2018, a program to study the
  spectrum and structure of $N^*$ states in various exclusive electroproduction channels using the new
  CLAS12 spectrometer commenced. These studies will probe the structure of $N^*$ states in the mass
  range up to $W$=3~GeV and for $Q^2$ as low as 0.05~GeV$^2$ and as high as 10-12~GeV$^2$, thus
  providing a means to access $N^*$ structure information spanning a broad range of distance scales.
  Quasi-real photoproduction studies are also planned to search for additional $N^*$ states, the so-called
  hybrid baryons, for which the glue serves as an active structural component. In this talk the $N^*$
  programs from both CLAS and CLAS12 will be reviewed.
\end{abstract}

% Head 1
\boldmath
\section{HALL B $N^*$ PROGRAM - OVERVIEW AND GOALS}
\unboldmath

The study of the spectrum and structure of excited nucleon states in Hall~B, referred to as the
$N^*$ program, represents one of the cornerstones of the physics program at Jefferson Laboratory
(JLab). The experimental program was carried out for $\sim$15~years using the large acceptance
CLAS spectrometer~\cite{mecking}. This system was decommissioned in 2012 and replaced with the
new large acceptance CLAS12 spectrometer~\cite{clas12} as part of the JLab 12~GeV upgrade project.
The program of CLAS12 includes a number of experiments as part of the continuing $N^*$ program in
Hall~B. The experiments will collect data over an unprecedented kinematic range for the study of nucleon
excited states. Data will be acquired in the range of $Q^2$ from 0.05 to 12~GeV$^2$, spanning the full
center-of-mass angular range of the decay final states. See Ref.~\cite{burkert-rev} for more details on
the overall CLAS12 physics program.

Both the CLAS $N^*$ program with electron and photon beams with energies up to 6~GeV (for
Ref.~\cite{burkert-fbs18} for a summary) and the CLAS12 $N^*$ program with electron beams up to 11~GeV
were designed to measure cross sections and spin observables over a broad kinematic range for a host of
different reaction channels. The main part of the $N^*$ program is based on electroproduction experiments
of various exclusive reactions employing an unpolarized liquid-hydrogen target. The central goals of this
focused program include:

\begin{itemize}
\item Unravel the spectrum of contributing excited nucleon states in a manner complementary to
  photoproduction data and confirm evidence for new baryon states with electroproduction data;
\item Explore the interplay between meson-baryon and dressed quark degrees of freedom in the
  structure of excited nucleon states via studies of the $Q^2$ evolution of the electro-excitation
  amplitudes;
\item Probe the momentum dependence of the dressed quark mass function from independent studies
  of the electro-excitation amplitudes of different excited nucleon states to shed light on the generation
  of $>$98\% of hadron mass;
\item Map out the momentum dependence of the dressed quark mass function at distances where the
  transition from the regime of quark-gluon confinement to pQCD is expected;
\item Understand the effect of non-point-like diquark correlations in explaining the structure of excited
  nucleon states of different quantum numbers via the electro-excitation amplitudes as a function of $Q^2$;
\item Search for predicted hybrid baryon states where glue is an active structural component.
\end{itemize}

The CLAS spectrometer was used to provide a significant fraction of the available world data for
the study of excited nucleon states with electron beams. To date nearly 50 papers have been published
to detail the measurement results specifically relevant for the $N^*$ program. The data include
differential cross section and various spin and polarization observables for the exclusive electroproduction
reactions $\gamma_v p \to  \pi^+ \pi^- p$, $\pi^0 p$, $ \pi^+ n$, $\eta p$, $K^+ \Lambda$, $K^+ \Sigma^0$,
$\omega p$, $\rho^0 p$, and $\phi p$. All data from CLAS are included in the CLAS Physics Database
\cite{physicsdb}. The analysis of data acquired from CLAS continues and additional analyses are currently in
progress~\cite{mokeev-fbs18}.

\section{EXTRACTION OF ELECTRO-EXCITATION AMPLITUDES}

The structure of excited nucleon states is more complex than what can be described accounting only for
quark degrees of freedom. At lower $Q^2$ ($\le$2~GeV$^2$), the structure of states is well described
by adding an external meson-baryon cloud to the underlying quark core. However, at higher $Q^2$
($\ge$3~GeV$^2$), the quark core is already the largest contributor to baryon structure. For
$Q^2 > 5$~GeV$^2$ we can probe the transition from the regime of confinement to perturbative Quantum
Chromodynamics (QCD). The detailed structure of the excited nucleon states is probed by the
$\gamma_v N N^*$ electro-excitation amplitudes as a function of $Q^2$. As the virtual photon has both
longitudinal and transverse polarization, these electro-excitation amplitudes (also called helicity amplitudes or
transition form factors) include both longitudinal ($S_{1/2}$) and transverse ($A_{1/2}$ and $A_{3/2}$ - 
where the subscripts refer to the helicity of the $\gamma_v p$ system) amplitudes. These amplitudes
directly reflect the underlying charge and current densities of these strongly interacting systems and elucidate
the relevant degrees of freedom of the excited nucleon states and their evolution with distance scale $Q^2$.
As $Q^2$ increases it is expected that these amplitudes will probe the coupling to the elementary quarks as
opposed to the dressed constituent quarks at lower $Q^2$. Such measurements hold promise to establish the
connection to QCD approaches in the strong interaction domain. Indeed, they represent our only source of
information on many facets of the non-perturbative strong interaction in the generation of $N^*$ states of
different quantum numbers from quarks and gluons. The data from CLAS with electron beam energies up to
6~GeV, spanned a range of $Q^2$ up to $\sim$5~GeV$^2$. With the coming data from CLAS12 with beam
energies up to 11~GeV, the $Q^2$ range of the data will extend to $\sim$12~GeV$^2$.

The extraction of the electrocouplings for most states up to $W \sim$1.8~GeV has been completed from CLAS
data based on exclusive analysis of the $\pi^0 p$, $\pi^+ n$, $\eta p$, and $\pi^+ \pi^- p$ reaction channels.
Table~\ref{listings} shows details on the different states for which the electrocouplings have been extracted
and the associated $Q^2$ range of the measurements~\cite{mokeev-web}. To complete these
extractions different analysis approaches have been employed. For the $\pi N$ and $\eta N$ channels the
electrocouplings have been extracted using either a unitary isobar model approach or via fixed-$t$ dispersion
relations. For the $\pi^+ \pi^- p$ channel, the data-driven JLab-Moscow State University model (JM) has been
employed. Ref.~\cite{aznauryan} contains a discussion of these different approaches.

%%%%%%%%%%%%%%%%%%%%%%%%%%%%%%%%%%%%%%%%%%%%%%%%%%%%%%%%%%%%%%
\begin{table}[htbp]
\caption{Excited nucleon states for which CLAS data have been used to extract the electro-excitation amplitudes.
The table shows the reaction channel, the states, and the $Q^2$ range of the data. See Ref.~\cite{mokeev-web}
for the electrocouplings and the associated references.}
\tabcolsep7pt\begin{tabular}{c|c|c} \hline
Reaction & $N^*$, $\Delta^*$ States & $Q^2$ Ranges of \\
Channel  &                                               & Electrocouplings (GeV$^2$) \\ \hline 
$\pi^0 p$, $\pi^+ n$ &  $\Delta(1232)3/2^+$                                                   & 0.16 - 6.0 \\
                                    & $N(1440)1/2^+$, $N(1520)3/2^-$, $N(1535)1/2^-$ & 0.30-4.16 \\ \hline
$\pi^+ n$                    & $N(1675)5/2^-$, $N(1680)5/2^+$, $N(1710)1/2^+$ & 1.6 - 4.5 \\ \hline
$\eta p$                     & $N(1535)1/2^-$                                                             & 0.2 - 2.9 \\ \hline
$\pi^+ \pi^- p$           & $N(1440)1/2^+$, $N(1520)3/2^-$                               & 0.25 - 1.5 \\
                                    & $\Delta(1620)1/2^-$, $N(1650)1/2^-$, $N(1680)5/2^+$ & 0.5 - 1.5 \\
                                    & $\Delta(1700)3/2^-$, $N(1720)3/2^+$, $N'(1720)3/2^+$ & \\ \hline
\end{tabular}
\label{listings}
\end{table}
%%%%%%%%%%%%%%%%%%%%%%%%%%%%%%%%%%%%%%%%%%%%%%%%%%%%%%%%%%%%%%

Over the past several years calculations of form factors and electro-excitation amplitudes with the
Dyson-Schwinger equation framework have shown that these fundamental quantities directly determined
from experimental data are intimately connected with the underlying momentum dependence of the dressed
quark mass function~\cite{mokeev-fbs18,roberts-fbs18}. Combining the data already produced by CLAS and
the data expected with the increased $Q^2$ reach of CLAS12 up to 12~GeV$^2$, a very broad range of quark
momenta will be spanned to further probe the dressed quark mass function moving from the regime of fully
dressed constituent quarks in the confinement regime toward the regime of only ``lightly'' dressed bare
quarks. In this manner, the data from the $N^*$ program in Hall~B can address critical and fundamental
questions on the nature of confinement and how the overwhelming majority of the visible mass in the universe
(i.e. the mass of the hadrons) is generated.

\boldmath
\section{IMPORTANCE OF $K^+ Y$ FINAL STATES}
\unboldmath

The majority of the advancements in understanding the spectrum and structure of excited nucleon states have
been provided by advanced analyses of the $\pi N$, $\pi \eta$, and $\pi^+ \pi^- p$ channels~\cite{dsc-fbs18}.
However, with the publication of the high statistics photoproduction $K^+ \Lambda$ and $K^+ \Sigma^0$ data
from CLAS, the potential and importance of the hyperon channels has been appreciated. In fact, the spectrum
of excited nucleon states listed in the recent edition of the Particle Data Group (PDG)~\cite{pdg-2018} has been
radically altered by the CLAS $K^+Y$ data, including not only the cross sections, but also the available
polarization observables. Table~\ref{nstar-evol} shows a comparison of the PDG evidence for a dozen $N^*$ and
$\Delta^*$ states (based on the PDG ``*'' rating) compared to just a decade ago. For most of these states the
$K^+Y$ data were a crucial input.

%%%%%%%%%%%%%%%%%%%%%%%%%%%%%%%%%%%%%%%%%%%%%%%%%%%%%%%%%%%%%%
\begin{table}[htbp]
  \caption{The evolution of our understanding of the excited nucleon spectrum. This table shows 12 states
    whose PDG ``assurance'' rating has increased over the past decade and the initial and/or final states where
    this evidence has come from. The $KY$ channels have been a critical aspect of this evolution for a number of
    states.}
\tabcolsep7pt\begin{tabular}{c|c|c|c|c|c|c} \hline
State               & PDG      & PDG  & $\pi N$ & $K\Lambda$ & $K \Sigma$ & $\gamma N$ \\
$N(mass)J^P$        & pre-2010 & 2018 &         &            &            &            \\ \hline
$N(1710)1/2^+$      & *** & **** & **** & ** & *  & **** \\ \hline
$N(1875)3/2^-$      &     & ***  & **   & *  & *  & **   \\ \hline
$N(1880)1/2^+$      &     & ***  & *    & ** & ** & **   \\ \hline
$N(1895)1/2^-$      &     & **** & *    & ** & ** & **** \\ \hline
$N(1900)3/2^+$      & **  & **** & **   & ** & ** & **** \\ \hline
$N(2000)5/2^+$      & *   & **   & *    &    &    & **   \\ \hline
$N(2100)1/2^+$      & *   & ***  & ***  & *  &    & **   \\ \hline
$N(2120)3/2^-$      &     & ***  & **   & ** & *  & ***  \\ \hline
$N(2060)5/2^-$      &     & ***  & **   & *  & *  & ***  \\ \hline
$\Delta(1600)3/2^+$ & *** & **** & ***  &    &    & **** \\ \hline
$\Delta(1900)1/2^-$ & **  & ***  & ***  &    & ** & ***  \\ \hline
$\Delta(2200)7/2^-$ & *   & ***  & **   &    & ** & ***  \\ \hline
\end{tabular}
\label{nstar-evol}
\end{table}
%%%%%%%%%%%%%%%%%%%%%%%%%%%%%%%%%%%%%%%%%%%%%%%%%%%%%%%%%%%%%%

It is also important to appreciate that the CLAS detector has provided the lion's share of the $K^+ \Lambda$
and $K^+ \Sigma^0$ world data for electroproduction in the nucleon resonance region ($W$ up to 2.5~GeV,
$Q^2$ from 0.3 to 5.4~GeV$^2$)~\cite{dsc-fbs16}. These data are available over a broad kinematic range and
have comparable or smaller uncertainties than for the often-used $\pi^+ \pi^- p$ electroproduction data from
CLAS. The available $KY$ data from CLAS include differential cross sections, separated structure functions,
and both recoil and beam-recoil transferred polarization. These $KY$ electroproduction data can be used to
confirm the signals of new baryon states observed in photoproduction in a complementary fashion. Within each
bin of $Q^2$, the determined states must have the same mass and decay widths. In such an analysis, the
electroproduction data can be used to verify the findings for the states shown in Table~\ref{nstar-evol}.

In order to extract the electrocouplings from the $KY$ data to gain insight and access into the structure of
the excited nucleon states, a suitable model of the reaction must be developed that accurately describes the
available data. Such an extraction model has not been sufficiently developed to employ in this regard. However,
there has been some recent advancement of a possible candidate isobar model that has been fit to the available
$K^+Y$ photo- and electroproduction data from CLAS and elsewhere~\cite{skoupil}. Another important factor
to consider is the fact that for higher-lying excited nucleon states where the decay strength to $\pi N$ is much
reduced compared to the dominant $\pi \pi N$ cross sections, the $KY$ channels will be critical to provide an
independent extraction of the electrocouplings for higher-lying excited nucleon states (that have decay strength
to both final states). Such comparisons are essential to provide confidence in the reliability of the results.

\boldmath
\section{CLAS12 $N^*$ PROGRAM OVERVIEW}
\unboldmath

The CLAS12 spectrometer in Hall~B was conceived, designed, and built as part of the recent JLab 12~GeV
upgrade project. The maximum beam energy in Hall~B is limited to 11~GeV. The nominal beam-target luminosity
for experiments with CLAS12 is $1\times 10^{35}$~cm$^{-2}$s$^{-1}$, a 10 times increase above the original
CLAS spectrometer. This is required in order to measure cross sections to higher $Q^2$.  CLAS12 was installed
in Hall~B in the period from 2012-2017. At the end of 2017 the detector system was commissioned during an
Engineering Run. The physics program officially began with data taking in Feb. 2018. Experiments with a
longitudinally polarized electron beam on both unpolarized liquid-hydrogen and liquid-deuterium targets have
collected significant amounts of data with more running planned on these targets in upcoming years. The $N^*$
program experiments with CLAS12 include:

\begin{itemize}
\item Nucleon Resonance Studies with CLAS12 - Studies of $N^*$, $\Delta^*$ spectrum and structure with
  a beam energy of 11~GeV focusing primarily on $\pi N$, $\eta N$, and $\pi^+ \pi^- p$ final states.
\item $KY$ Electroproduction with CLAS12 - Studies of  $N^*$, $\Delta^*$ spectrum and structure with
  a beam energy of 11~GeV focusing primarily on $K^+ \Lambda$ and $K^+ \Sigma^0$ final states.
\item $N^*$ Studies via $KY$ Electroproduction at 6.6 and 8.8~GeV - Extension of experiment at 11~GeV
  at lower beam energies to provide precision data at $Q^2$ overlapping existing data from CLAS.
\item A Search for Hybrid Baryons in Hall~B with CLAS12 - A search for baryon states with glue as an
  active structure component predicted by lattice QCD calculations. This experiment will use beam energies
  of 6.6 and 8.8~GeV and focus on measurements of electrocouplings at $Q^2 < 0.5$~GeV$^2$ where
  the amplitudes are predicted to have a significantly different $Q^2$ dependence compared to three quark
  baryons.
\end{itemize}

At the current time the CLAS Collaboration is gearing up to begin analysis of the significant amount of data
already collected with CLAS12. To date this amounts to $\sim$400~mC of collected charge and $\sim$3~PB
of data to tape storage after roughly of year of data taking. In order to be ready for physics analysis, a
significant level of effort has been being geared toward optimizing detector alignment in both the Forward and
Central Detectors of CLAS12 using data collected at zero field, quantifying our knowledge of tracking efficiency
and improving algorithms to maximize this efficiency, understanding the momentum distortions to enable effective
corrections to the reconstructions to optimize missing mass resolutions for identification of exclusive final states
and minimizing backgrounds, and developing all corrections necessary to enable optimal timing resolution to enable
the best possible charged particle identification versus momentum.

The initial analyses and publications based on data collected with CLAS12 will involve yield ratios from
extractions of various beam spin asymmetries where a detailed understanding of detector acceptance
and efficiencies are not essential. The important tasks that lie ahead are understanding the systematics
of yield normalization for cross section extractions, to develop accurate kinematic corrections to optimize
momentum and missing mass resolutions due to undesired effects due to limitations of our knowledge of
magnetic field, detector geometry, alignment, and reconstruction biases. A key part of our efforts to
extract cross sections from CLAS12 will also involve development of accurate radiative correction models
to account for important bin migration effects over our full kinematic phase space for our inclusive,
semi-inclusive, and exclusive reactions.

\section{CONCLUDING REMARKS}

The study of the spectrum and structure of excited nucleon states in Hall~B at Jefferson Laboratory,
referred to as the $N^*$ program, is based on three different but equally essential aspects. The
{\em first} is providing experimental data of high quality and broad kinematic coverage for a number of
different exclusive reaction channels. The observables are in the form of differential cross sections,
separated structure functions, beam spin asymmetries, and polarization observables span invariant energies
up to 4~GeV, four-momentum transfers $Q^2$ based on data from CLAS up to 5~GeV$^2$ with electron
beams up to 6~GeV and on data from CLAS12 up to $Q^2$ of 12~GeV$^2$. With the large acceptance
CLAS and CLAS12 spectrometers, these data span the full angular phase space in the center-of-mass
system for the final state reaction products. To date, the data from CLAS in the nucleon resonance region
for both photoproduction and electroproduction experiments dominate the world database. In the coming
years the data measurements from CLAS12 will significantly overlap and extend these data.

The {\em second} essential aspect of the Hall~B $N^*$ program is to facilitate the development of reaction
models that accurately describe the available experimental data. Such models are essential tools in order to
be able to extract the electrocouplings of the contributing $N^*$ and $\Delta^*$ states. The {\em third}
essential aspect is the development of the QCD-rooted approaches in order to connect the data to theory to
make progress in understanding strong interaction dynamics of dressed quarks and their confinement in baryons
over a broad range of $Q^2$. High quality experimental data are essential to address the most challenging
problems of the Standard Model of fundamental particles and interactions on the nature of hadron mass,
confinement, and the emergence of excited nucleon states of different quantum numbers from QCD.

\section{ACKNOWLEDGMENTS}

The author thanks his colleague Victor Mokeev for many useful discussions and insights on the topics
discussed in this presentation. This material is based upon work supported by the U.S. Department of
Energy, Office of Science, Office of Nuclear Physics under contract DE-AC05-06OR23177.

\end{document}